\documentclass[accepted]{uai2024} % after acceptance, for a revised version; 
\usepackage{amsmath,amsfonts,bm}

\def\eqref#1{equation~\ref{#1}}
\def\1{\bm{1}}

\DeclareMathAlphabet{\mathsfit}{\encodingdefault}{\sfdefault}{m}{sl}
\SetMathAlphabet{\mathsfit}{bold}{\encodingdefault}{\sfdefault}{bx}{n}

\def\gD{{\mathcal{D}}}

\def\gH{{\mathcal{H}}}
\def\gI{{\mathcal{I}}}

\def\gL{{\mathcal{L}}}

\def\gN{{\mathcal{N}}}

\def\gU{{\mathcal{U}}}

\DeclareMathOperator*{\argmax}{arg\,max}
\DeclareMathOperator*{\argmin}{arg\,min}

\usepackage[american]{babel}
\usepackage{graphicx}
\usepackage{subfig}
\usepackage{natbib} % has a nice set of citation styles and commands
\usepackage{mathtools} % amsmath with fixes and additions
\usepackage{booktabs} % commands to create good-looking tables
\usepackage{tikz} % nice language for creating drawings and diagrams
\usepackage{multirow,tabularx}
\usepackage[linesnumbered,ruled,vlined]{algorithm2e}

\SetCommentSty{mycommfont}
\SetKwInput{KwInput}{Input}
\SetKwInput{KwOutput}{Output}

\usepackage[capitalize,noabbrev]{cleveref}

\usepackage[textsize=tiny]{todonotes}
\title{Language-Model Prior Overcomes Cold-Start Items}

\author[1]{Shiyu Wang\thanks{Work done during the author’s internship at AWS AI Labs.}}
\author[2]{Hao Ding}
\author[2]{Yupeng Gu}
\author[2]{Sergul Aydore}
\author[2]{Kousha Kalantari}
\author[2]{Branislav Kveton\thanks{Corresponding author: kveton@adobe.com}}
\affil[1]{%
    Salesforce Research\\
    Palo Alto, California, USA
}
\affil[2]{%
    AWS AI Labs\\
    Santa Clara, California, USA
}

\begin{document}
\maketitle
\begin{abstract}
The growth of recommender systems (RecSys) is driven by digitization and the need for personalized content in areas such as e-commerce and video streaming. The content in these systems often changes rapidly and therefore they constantly face the ongoing cold-start problem, where new items lack interaction data and are hard to value. Existing solutions for the cold-start problem, such as content-based recommenders and hybrid methods, leverage item metadata to determine item similarities. The main challenge with these methods is their reliance on structured and informative metadata to capture detailed item similarities, which may not always be available. This paper introduces a novel approach for cold-start item recommendation that utilizes the language model (LM) to estimate item similarities, which are further integrated as a Bayesian prior with classic recommender systems. This approach is generic and able to boost the performance of various recommenders. Specifically, our experiments integrate it with both sequential and collaborative filtering-based recommender and evaluate it on two real-world datasets, demonstrating the enhanced performance of the proposed approach. Code can be found at \url{https://github.com/awslabs/language-model-prior-4-item-cold-start}.
\end{abstract}

\section{Introduction}\label{sec:intro}
% Recommendation systems are very import.
% Cold-start item recommendation. Why this problem arises?
\begin{figure}[htb!]
\begin{center}
\includegraphics[width=0.3\textwidth]{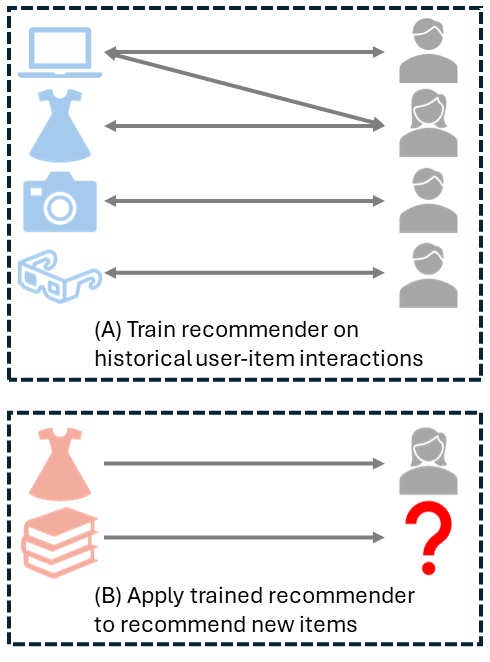}
\caption{Recommender is trained on historical user-item interactions, and then used to recommend new items, including those that previously appeared (i.e., dress) and newly introduced cold-start items (i.e., book).}
\label{fig:rec}
\end{center}
\vspace{-10mm}
\end{figure}
The field of recommender systems (RecSys) has witnessed tremendous growth over the last few years~\citep{sasrec,hrnn,zesrec,lmrec,prerec,textrec,trendrec}, driven by the increasing service digitization and the rising demand for personalized content across diverse platforms such as e-commerce and video streaming. Despite significant advancements, the unresolved item cold start problem remains a critical challenge. It arises with newly introduced items lacking sufficient interaction data, and thus struggling to be accurately recommended to users. For instance, Figure~\ref{fig:rec} illustrates that items such as laptop, dress, camera and glasses that have appeared in historical data are typically easier to recommend, whereas those never-before-seen items are challenging. This issue is especially severe in dynamic environments, such as news recommendation, where new items are constantly introduced, making it tough to identify similarities among cold items due to insufficient information.

% Two major solutions to cold-start item recommendation: (1) content-based recommendation and (2) other hybrid methods.
To address the item cold start problem, previous works primarily fall into two major categories: content-based recommendation and hybrid methods. Both approaches focus on leveraging additional item metadata to uncover item similarities. Content-based recommendation approaches such as~\citep{oord13deep,volkovs17contentbased,volkovs17dropoutnet} tackle the issue by utilizing item metadata such as item category. These methods analyze the item content and recommend similar items by aligning them with user preferences, rather than relying heavily on past user-item interactions. For example, a model might recommend a new fantasy novel to a fan of the genre, despite limited interaction data on the novel itself. Another line of works focuses on hybrid methods~\citep{wang11collaborative,zhang16collaborative,pan2019warm,zhou20s3rec,liu21augmenting,han22addressing}, which combine the strength of both content-based and collaborative filtering (CF) techniques. These approaches integrate user behavior with item attributes to generate recommendations, aiming to capitalize on both aspects. Hybrid methods are particularly notable for delivering precise recommendations by encompassing a wide spectrum of user preferences and item features. Overall, both strategies aim to tap into the item metadata to provide prior knowledge about item similarities. However, the main problem with these methods is twofold: (1) they require structured metadata, which is not always available, and (2) the structured metadata may be uninformative and unable to capture fine-grained item similarities. For example, consider a scenario with only five unique categories for all items, with a skewed distribution where the largest category comprises 80\% of the items.

% Our method and why it stands out: (1) we proposed a very general regression with a Bayesian regularizer. we can incorporate the prior information in a very simple way by taking into account similarity among items. It can be applied to both settings: transformer and Bayesian regression.
Recent advancements in Language Models (LMs) enable extracting insights from unstructured textual metadata, like product descriptions or movie synopses, using pre-trained models such as BERT~\citep{bert} and Falcon~\citep{falcon}.. This approach leverages the inherent prior knowledge of LMs to uncover item similarities, even with limited available interactions or structured item metadata. Existing works~\citep{zesrec} focus on generating text embedding from the pre-trained LMs based on the item textual metadata. However, directly using LMs embedding of items as the input of recommender may introduce a vast amount of information, not all of which is relevant to the recommendation task at hand. Also, it can significantly increase the dimensionality of data the recommender should process, and limits the flexibility of how the item meta information is integrated into the recommendation system.

Therefore, in contrast to previous works, this paper investigates the possibilities and challenges of implicitly harnessing LMs to inject prior knowledge of items. Specifically, we propose a framework that integrates a Bayesian regularizer into the training process of the RecSys. The regularizer takes into account the semantic similarity among items, leveraging LM-encoded prior knowledge to learn fine-grained item embeddings in the continuous latent space. The proposed approach is generic and can be adopted in a plug-and-play fashion in any sequential~\citep{sasrec,hrnn} or CF-based~\citep{bpr,cdl} RecSys. Our contributions can be summarized as follows:
\begin{itemize}[leftmargin=15pt]
\item We introduce a novel Bayesian framework that leverages LMs embeddings of item metadata as the prior to address the item cold start problem in RecSys. This framework leverages the rich semantic information from LMs to improve the ability of the recommender to understand and recommend new items effectively.
\item The proposed Bayesian prior is integrated into RecSys as a regularizer, and therefore is generic and able to enhance various recommenders by supplementing their learning objectives.
\item We evaluated the proposed method using both sequential (i.e., SASRec) and CF-based (i.e., BPRMF) recommenders on two real-world datasets from distinct domains. The empirical results demonstrate the enhanced performance of the proposed approach, which improves SASRec by 17.78$\%$ regarding normalized discounted cumulative gain.
\end{itemize} 

% \todob{Be more concrete in contributions. To showcase the generality of our approach, we apply it to two models and validate it on two datasets.}

\section{Related Work}
As collaborative filtering-based recommenders~\citep{schafer2007collaborative, koren2021advances, chen2018survey} learn user and item embeddings based on their interactions, they cannot deal with cold-start items without using their metadata. In general, recommenders that are capable of recommending cold-start items are classified into two scenarios: (1) content-based cold-start item recommendation and (2) hybrid methods, which are introduced as follows.

\subsection{Content-based cold-start item recommendation}
Content-based recommenders start with user and item features, and therefore are suitable and play a major role for recommending cold-start items. User-item interactions can be incorporated as additional features to train the recommender. This process does not intend to learn item embeddings but, instead, leverage item features to during training process. This is initially presented as \citet{volkovs17contentbased} and later published as \citet{volkovs17dropoutnet}. A deep content-based recommender that performs well on cold-start items is proposed by \citet{oord13deep}. While content-based recommenders that rely on item features can be effective, it learns the residual distribution of item characteristics based solely on their features, and therefore often misses the nuanced, implicit relationships between items that can usually be captured through continuous item embeddings. Content-based recommenders typically have limited generalization capabilities compared to embedding-based methods. The latter can better generalize to new or unseen items by discerning their position in the embedding space relative to other items, enhancing recommendation quality.

\subsection{Hybrid method for cold-start item recommendation}
% Collaborative filtering recommenders, which learn user and item embeddings, need user-item interactions and thus cannot deal with cold items.
Hybrid approaches combine different strategies to better capture similarity among items. \citet{wang11collaborative} proposed collaborative topic regression. The approach combines probabilistic matrix factorization with a topic model prior and learns the distribution over the topics, which are over item descriptions. The prior is applied as a regularizer in regression and yields a classic Bayesian recommendation method. This method does not learn the user-item interactions in a sequential way and therefore might suffer from capturing dependencies among items compared with sequential-based recommenders. Other methods may learn the item embeddings to uncover their similarities in the latent space. For instance, \cite{zhang16collaborative} jointly learned the latent representations in collaborative filtering as well as semantic representations of items. The semantic representations are extracted from structural, textual, and visual content. \cite{liu21augmenting} addressed the poor performance of transformers on short sequences. The key idea is to extend these sequences by simulated prior items, which are added before the actual items. \citet{zhou20s3rec} proposed a transformer that takes into account various forms of correlation between context and sequential data via the attention mechanism. However, this approach is tied to the framework of transformer and lacks generalization to other recommenders. \citet{han22addressing} used empirical Bayes to estimate the prior distribution of engagement metrics of an item conditioned on non-behavioral signals. The engagement with the item can be then incorporated using posterior updates. We propose a generic and novel Bayesian-based approach that can be applied to various recommenders, ranging from sequential-based approach to CF-based recommenders, and achieves enhanced performance compared to the corresponding original method. 

% \sergul{It is still unclear to me in what aspects your approach is different or brings value on top of above prior work.}

\section{Problem Formulation}

We consider a recommendation problem where we recommend items to users. Let $\gI$ be the set of $N$ recommended items, indexed by $i \in \gI$. Let $\gU$ be the set of $M$ users, indexed by $j \in \gU$. We denote by $\gD \subset \gI \times \gU \times \mathbb{N}$ a dataset of user-item interactions that are indexed by time. Specifically, each interaction $(i, j, t) \in \gD$ is a triplet of item $i$, user $j$, and time $t$. Without loss of generality, and since the time of making a recommendation is not used in our models, we assume that the time is represented by positive integers that index the recommendations. The first recommended item for each user is at time $t = 1$. If the user is recommended an item at time $t > 1$, they must have been recommended an item at time $t - 1$ as well. We let:
\begin{align*}
  \gH_j
  = \{i \in \gI: (i, j ,t) \in \gD\}
\end{align*}
be the set of items that user $j$ interacted with and:
\begin{align*}
  \gH_{j, t}
  = \{i \in \gI: (i, j, \ell) \in \gD, \, \ell < t\}
\end{align*}
be the set of items that user $j$ interacted with before time $t$. We denote by $I_{j, t}$ the item that user $j$ interacted with at time $t$. Each item $i$ is associated with metadata $E_i$ that describes it. If the item is a movie, its metadata could be its title, director, or genre. If the item is a product, its metadata could be its name, price, and reviews.

Our objective is to learn to recommend item $I_{j, t}$, that user $j$ interacts with at time $t$, given the history of their past interactions. This item is a random variable:
\begin{align}
  I_{j, t}
  \sim p(\cdot \mid \gH_{j, t}, \{E_i\}_{i \in \gI})\,,
  \label{eq:objective}
\end{align}
where $p(\cdot \mid \gH_{j, t}, \{E_i\}_{i \in \gI})$ is a distribution over all items $\gI$ conditioned on the history of user $j$ up to time $t$, $\gH_{j, t}$, and the metadata of all items, $\{E_i\}_{i \in \gI}$. The dependence on all metadata allows us to model that the probability of recommending an item could depend on its metadata. Our goal is to learn the conditional distribution $p$ in \eqref{eq:objective}.

\begin{figure}[t]
\begin{center}
\includegraphics[width=0.4\textwidth]{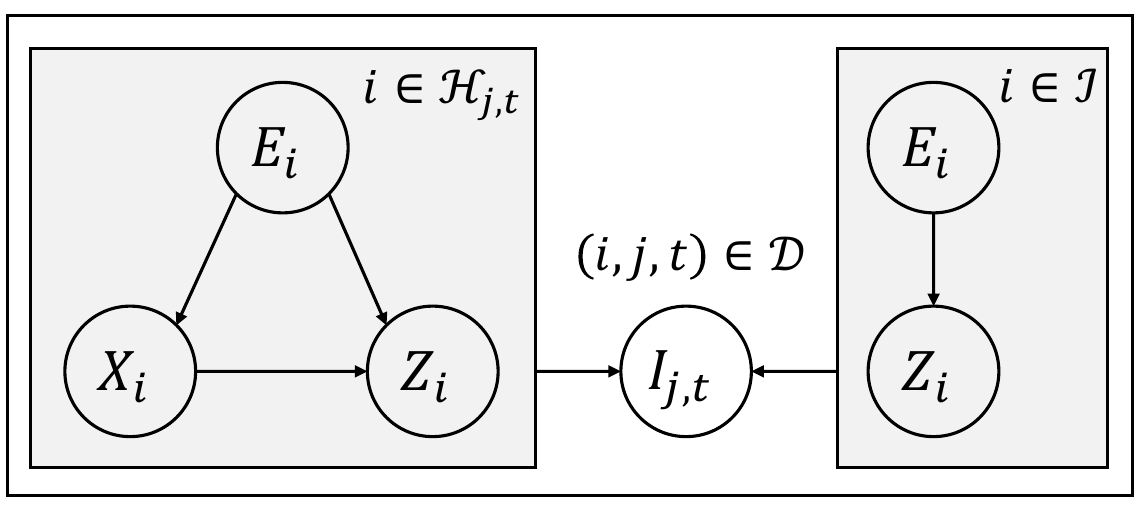}
\caption{The probabilistic model of our proposed method. Overall, $E_i$ is the metadata of item $i$, $X_i$ is the LMs embedding encoded from $E_i$ serving as the prior knowledge when learning $Z_i$, and $Z_i$ is the learned item embedding by the recommender from $E_i$ and $X_i$. To recommend the item at time step $t$, we leverage all items that have interacted with the user $j$ to learn a distribution over all items in $\gI$, among which we recommend the one with the highest likelihood.}
\label{fig:prob}
\end{center}
\end{figure}

\section{Sequential Recommender}
\label{sec:sequential recommender}

This section has two main parts. In \cref{sec:algorithm}, we introduce a general sequential recommender. In \cref{sec:regularized loss}, we present a regularized variant of its learning objective, where the regularizer can be interpreted as a Bayesian prior probability. Finally, in \cref{sec:more informative prior}, we present a more informative prior based on language models. This approach, while showcased with sequential recommenders, can be easily adapted to other recommenders, which we show in \cref{sec:experiments}.

\subsection{Algorithm}
\label{sec:algorithm}

We take the following approach. Since item metatada $E_i$ can have many forms, such as an image of the item, its text description, or reviews, we encode them into a vector $Z_i \in \mathbb{R}^d$. Specifically, we let $Z_i = g_\gamma(E_i)$, where $g_\gamma$ is a metadata encoder parameterized by $\gamma$. With this in mind, we approximate the conditional distribution $p$ in \eqref{eq:objective} using a function:
\begin{align}
  f_\theta(\cdot \mid \gH_{j, t}, \{g_\gamma(E_i)\}_{i \in \gI})
  \label{eq:model}
\end{align}
parameterized by $\theta$. This function can be a recurrent neural network \citep{hidasi16sessionbased} or a transformer \citep{kang18selfattentive}, and its output is a distribution over all items $\gI$. The function depends on the history of user $j$ up to time $t$, $\gH_{j, t}$, and the encoded metadata of all items, $\{g_\gamma(E_i)\}_{i \in \gI}$.

Learning of $f_\theta$ is typically formalized as maximizing the probability of making correct recommendations over the whole dataset:
\begin{align*}
  \argmax_{\theta, \gamma} \prod_{(i, j, t) \in \gD}
  f_\theta(i \mid \gH_{j, t}, \{g_\gamma(E_i)\}_{i \in \gI})\,.
\end{align*}
This can be also viewed as minimizing the cross-entropy loss, $\argmin_{\theta, \gamma} \gL(\gD; \theta, \gamma)$, where the cross-entropy loss is defined as:
\begin{align}
  \gL(\gD; \theta, \gamma)
  = \sum_{(i, j, t) \in \gD}
  - \log f_\theta(i \mid \gH_{j, t}, \{g_\gamma(E_i)\}_{i \in \gI})\,.
  \label{eq:cross-entropy loss}
\end{align}
This loss can be minimized using existing optimizers, such as stochastic gradient descent \citep{zinkevich03online} and Adam \citep{kingma15adam}. The graphical model of our approach is illustrated in \cref{fig:prob}.

\subsection{Regularized Loss}
\label{sec:regularized loss}

In practice, a regularized loss is optimized instead, which can be written as:
\begin{align}
  \hat{\gL}(\gD; \theta, \gamma, \rho)
  = \gL(\gD; \theta, \gamma) + \rho \sum_{i \in \gI} \|Z_i\|_2^2\,.
  \label{eq:regularized loss}
\end{align}
Here $\rho > 0$ is a regularization parameter that determines the strength of the regularization. The regularization term can be interpreted as a prior probability in a Bayesian formulation as follows. First, fix item $i$ and note that:
\begin{align*}
  \|Z_i\|_2^2
  & = - 2 \log \exp\left[- \frac{1}{2} \|Z_i\|_2^2\right] \\
  & = - 2 \log \exp\left[- \frac{1}{2} (Z_i - \mathbf{0}_d)^\top I_d
  (Z_i - \mathbf{0}_d)\right]\,.
\end{align*}
Therefore, up to normalizing constants, $\|Z_i\|_2^2$ is proportional to the logarithm of the probability density function (PDF) of $\gN(Z_i; \mathbf{0}_d, I_d)$. Following the same line of reasoning, we can derive the following:
\begin{align*}
  & \sum_{i \in \gI} \|Z_i\|_2^2 = - 2 \log \exp\left[- \frac{1}{2} \sum_{i \in \gI} \|Z_i\|_2^2\right] \\
  & \quad = - 2 \log \exp\left[- \frac{1}{2} (Z - \mathbf{0}_{d N})^\top I_{d N}
  (Z - \mathbf{0}_{d N})\right]\,,
\end{align*}
where $Z = (Z_i)_{i \in \gI}$ is the concatenation of all item embeddings. Therefore, up to normalizing constants, $\sum_{i \in \gI} \|Z_i\|_2^2$ is proportional to the logarithm of the PDF of $\gN(Z; \mathbf{0}_{d N}, I_{d N})$.

To complete our argument, we note the following. First, the cross-entropy loss $\gL(\gD; \theta, \gamma)$ in \eqref{eq:regularized loss} is the negative log-likelihood of data $\gD$ given learned model parameters, including learned item embeddings $Z$. Second, the regularizer $\sum_{i \in \gI} \|Z_i\|_2^2$ in \eqref{eq:regularized loss} is proportional to the logarithm of the probability that item embeddings are $Z$. It follows that \eqref{eq:regularized loss} is the posterior probability of model parameters, including learned item embeddings $Z$, and thus $\sum_{i \in \gI} \|Z_i\|_2^2$ is proportional to the logarithm of the prior probability of $Z$.

\section{A More Informative Prior}
\label{sec:more informative prior}

The regularizer in \eqref{eq:regularized loss} is not informative. In particular, it is proportional to the logarithm of a probability where item embeddings $Z_i$ do not depend on each other. They are simply centered at zero vectors and have unit covariances.

In this work, we replace \eqref{eq:regularized loss} with:
\begin{align}
  \hat{\gL}(\gD; \theta, \gamma, \rho)
  = \gL(\gD; \theta, \gamma) + \rho \sum_{i, k \in \gI} s_{i, k} \|Z_i - Z_k\|_2^2\,,
  \label{eq:obj_reg_local}
\end{align}
where $s_{i, k} \in [0, 1]$ it the \emph{prior similarity} of items $i$ and $k$, and $\rho > 0$ is a regularization parameter. We assume that the similarity is symmetric, $s_{i, k} = s_{k, i}$ for any $i, k \in \gI$.

As we show next, the new regularizer encodes the similarities of items. Specifically, it can be viewed as a Bayesian prior probability over their embeddings. To see this, note that:
\begin{align*}
  \|Z_i - Z_k\|_2^2
  = Z_i^\top Z_i - Z_i^\top Z_k - Z_k^\top Z_i + Z_k^\top Z_k\,.
\end{align*}
Thus:
\begin{align*}
  & \sum_{i, k \in \gI} s_{i, k} \|Z_i - Z_k\|_2^2 \\
  & = - 2 \log \exp\left[- \frac{1}{2} \sum_{i, k \in \gI} s_{i, k} \|Z_i - Z_k\|_2^2\right] \\
  & = - 2 \log \exp\left[- \frac{1}{2} \left[\sum_{i \neq k} \lambda_{i, k} Z_i^\top Z_k +
  \sum_{i \in \gI} \lambda_i Z_i^\top Z_i\right]\right]\,,
\end{align*}
where:
\begin{align*}
  \forall i \neq k: \lambda_{i, k}
  = - 2 s_{i, k}\,, \quad
  \forall i \in \gI: \lambda_i
  = 2 \sum_{k \neq i} s_{i, k}\,.
\end{align*}
We can further rewrite this as a single prior distribution as follows. Let $\Lambda = (\Lambda_{i, k})_{i, k \in \gI}$ be a $d N \times d N$ block matrix, where each block $\Lambda_{i, k}$ is a $d \times d$ matrix. Let:
\begin{align*}
  \forall i \neq k: \Lambda_{i, k}
  = \lambda_{i, k} I_d\,, \quad
  \forall i \in \gI: \Lambda_{i, i}
  = \lambda_i I_d\,.
\end{align*}
Then $\sum_{i, k \in \gI} s_{i, k} \|Z_i - Z_k\|_2^2 =$
\begin{align*}
  - 2 \log \exp\left[- \frac{1}{2} (Z - \mathbf{0}_{d N})^\top \Lambda
  (Z - \mathbf{0}_{d N})\right]\,,
\end{align*}
The new learning objective in \eqref{eq:obj_reg_local} is very different from \eqref{eq:regularized loss}. The latter treats items independently, because each is regularized independently using $\|Z_i\|_2$,  while the former encodes dependencies among the items, because of $\|Z_i - Z_k\|_2$. The regularizer in \eqref{eq:obj_reg_local} can also be viewed as a form of graph-based regularization \citep{belkin04regularization}, where the items are nodes and their similarities are edge weights.

\section{Learning Prior}

We compute the similarities $s_{i, k}$ using pre-trained \emph{language models (LMs)}. Let $X_i = \mathrm{LM}(E_i)\in\mathbb R^{d'}$ be the LM embedding of metadata of item $i$. Then $s_{i, k} = g(X_i, X_k)$ for some function $g$. We make two assumptions on function $g$. The first assumption is non-negativity: for any $X_i$ and $X_k$, $g(X_i, X_k) \geq 0$. The second assumption is symmetry: for any $X_i$ and $X_k$, $g(X_i, X_k) = g(X_k, X_i)$.

Specifically, we define the similarity of items $i$ and $k$ as:
\begin{align}
  s_{i, k}
  = g(X_i, X_k)
  = \exp\left[- \frac{\lambda}{2} \|X_i - X_k\|_2^2\right]\,,
  \label{eq:s}
\end{align}
where $\lambda > 0$ is a tunable parameter. We set the parameter $\lambda$ as follows. Let $\hat{\mu} = \frac{1}{N} \sum_{i \in \gI} X_i$ be the mean embedding across all items. Let:
\begin{align}
  \hat{\sigma}^2
  = \frac{1}{Nd'} \sum_{i \in \gI} (X_i - \hat{\mu})^\top(X_i - \hat{\mu})
\end{align}
be the average variance, across all items and embedding dimensions. Then $\lambda = 1 / \hat{\sigma}^2$. In an essence, $1 / \lambda$ is one standard deviation defined by data, across all items and embedding dimensions, akin to that in a normal distribution.

To learn more fine-grained item similarities, we also consider estimating a separate similarity kernel for each item $i$, which takes the local geometry of the language-model embedding space into account. Specifically, we assume that $X_i\sim\gN(\mu_i, \Sigma_i)$, where $\mu_i$ and $\Sigma_i$ can be estimated empirically by $X_i$ and its neighboring item embeddings. Specifically, for any neighboring item $i$ and item $k$, $X_i-X_k\sim\gN(\bm 0, 2\Sigma_i)$. Let $\gN_i$ be indices of the $K$-nearest neighbors of embedding $X_i$ (including $X_i$). Then \eqref{eq:s} becomes: 
% \todob{Use \textbackslash eqref when referring to equations.}
\begin{align}
  s_{i, k}
  = \exp\left[- \frac{1}{2} (X_i - X_k)^\top \hat{\Sigma}_i^{-1} (X_i - X_k)\right]\,,
  \label{eq:s_emp}
\end{align}
where
\begin{align}
  \hat{\mu}_i
  & = \frac{1}{K} \sum_{\ell \in \gN_i} X_\ell \nonumber\\
  \hat{\Sigma}_i
  & = \frac{1}{K} \sum_{\ell \in \gN_i} (X_\ell - \hat{\mu}_i) (X_\ell - \hat{\mu}_i)^\top\,.
  \label{eq:local}
\end{align}
The training process is further described in \cref{alg:train}.

\begin{algorithm}
\scriptsize
\DontPrintSemicolon
\KwInput{A set of historical data of $M$ users: $\gH_j\in\gD$}
\KwInput{A set of embeddings of all items: $X_1, X_2, ..., X_N$}
\KwInput{Pre-defined hyperparameter $\rho$ and $K$}
\While{Converge}
   {
   \For{a batch of $B$ users and their interactions with items sampled from $\gD$}{
   \For{each item $i$ interacted with user $j$ in the batch}{
   Encode $i$ by $Z_i=g_\gamma(E_i)$ as in \eqref{eq:model} \\
    Identify its $K$ nearest items in $\gN_i$\\
   Estimate $\hat{\mu}_i$ and $\hat{\Sigma}_i$ as in \eqref{eq:local}\\
   Compute similarity among $i$ and $\gN_i$ as in \eqref{eq:s_emp}
   }
   Compute recommendation loss in \eqref{eq:cross-entropy loss} based on the chosen recommender\\
   Compute graph regularization term as in \eqref{eq:obj_reg_local}\\
   Compute the loss $\gL$ as in \eqref{eq:obj_reg_local}\\
   Update model parameters by gradient descent on $\gL$
   }
   }
\caption{Training process}
\label{alg:train}
\end{algorithm}

Note that identifying $K$ nearest items can be conducted offline, which can largely reduce the computational burden during training.

\section{Experiments}
\label{sec:experiments}

In this section, we start with introducing two real-world datasets to evaluate our proposed method. Next, we present a set of comparison baselines employed in the experiments. Following that, we delve into the explanation of the chosen evaluation metrics in the assessment process. Finally, we evaluate the proposed method from two distinct perspectives: (1) examining the efficacy of incorporating graph regularization, and (2) analyzing the influence of the penalty associated with the graph regularization component.
\begin{figure*}[hbt!]
\centering
\subfloat{\includegraphics[width=0.25\textwidth]{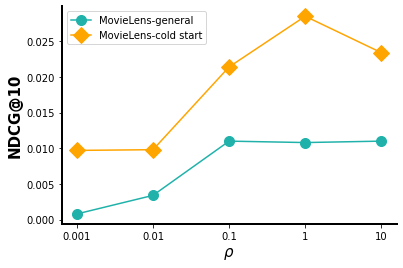}}\hfil
\subfloat{\includegraphics[width=0.25\textwidth]{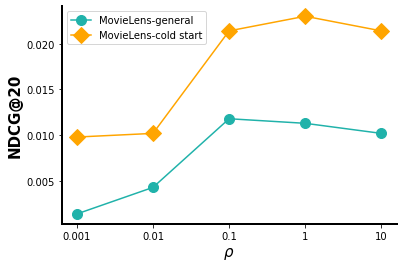}}\hfil 
\subfloat{\includegraphics[width=0.25\textwidth]{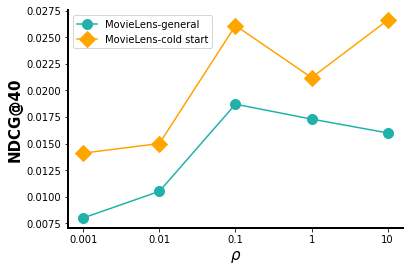}} 

\subfloat{\includegraphics[width=0.25\textwidth]{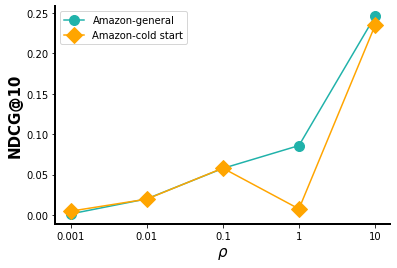}}\hfil   
\subfloat{\includegraphics[width=0.25\textwidth]{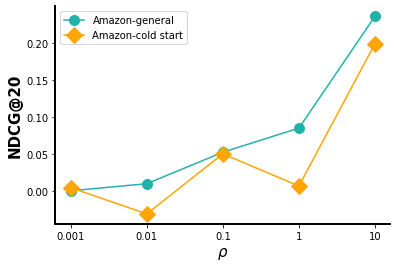}}\hfil
\subfloat{\includegraphics[width=0.25\textwidth]{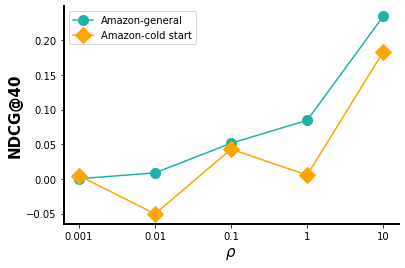}}
\caption{Tuning hyperparameter $\rho$: overall performance on testing set of SASRec-ours regarding Normalized Discounted Cumulative Gain (NDCG).}\label{fig:ndcg}
\vspace{-7mm}
\end{figure*}

\begin{figure*}[hbt!]
\centering
\subfloat{\includegraphics[width=0.25\textwidth]{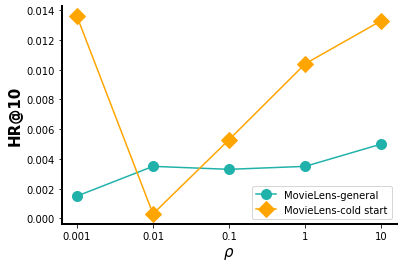}}\hfil
\subfloat{\includegraphics[width=0.25\textwidth]{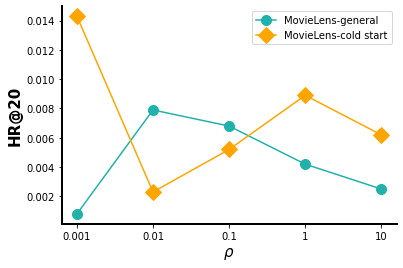}}\hfil 
\subfloat{\includegraphics[width=0.25\textwidth]{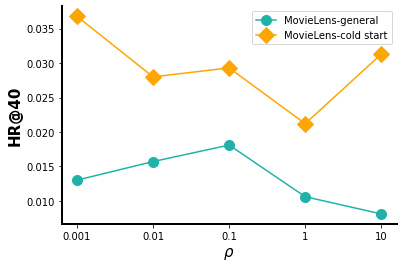}} 

\subfloat{\includegraphics[width=0.25\textwidth]{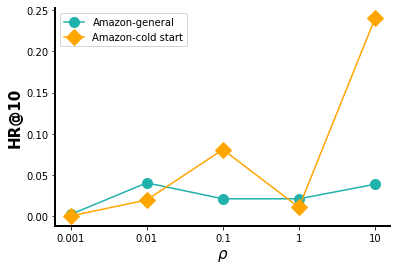}}\hfil   
\subfloat{\includegraphics[width=0.25\textwidth]{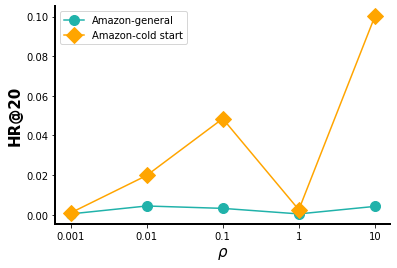}}\hfil
\subfloat{\includegraphics[width=0.25\textwidth]{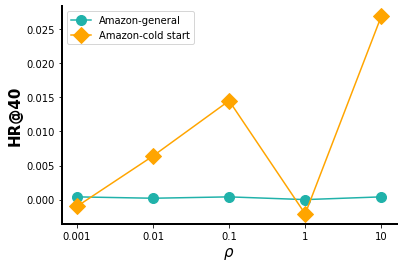}}
\caption{Tuning hyperparameter $\rho$: overall performance on testing set of SASRec-ours regarding Hit Ratio (HR).}\label{fig:hr}
\vspace{-5mm}
\end{figure*}

\subsection{Datasets}
We employ two real-world datasets, MovieLens (i.e., MovieLens 25M) and Amazon (i.e., Amazon Prime Pantry 5-core), to evaluate the proposed method. Additionally, let cold-start (CS) items be items that have at most five instances in the whole dataset, and we further define cold-start (CS) users as those interacted with at least one cold-start item. Users interacted with cold-start items are extracted from each dataset to assess the effectiveness of the proposed method on recommending cold-start items. Specifically, details regarding two datasets are presented in Table~\ref{tab:data}.

\begin{table}[]
 \caption{Details of two datasets used to evaluate the proposed method.}
\vspace{-3mm}
\begin{adjustbox}{width=0.3\textwidth,center}
\begin{tabular}{ccc}
\hline
Datasets                & MovieLens & Amazon \\ \hline
No. of users            &  162,541             &  14,180                           \\ \hline
No. of items            &  59,047             &  4,968                           \\ \hline
No. of CS users &   6,063            &           1,936                  \\ \hline
No. of CS items &   28,840            &           446                  \\ \hline
\end{tabular}
\end{adjustbox}
\label{tab:data}
\vspace{-5mm}
\end{table}

\textbf{MovieLens} We collect MovieLens 25 dataset from grouplens\footnote{https://grouplens.org/datasets/movielens/25m/} that contains 25 million movie ratings applied to 162,541 users. A sequence of types of movies interacted with a user is treated as a sample.

\textbf{Amazon} We employ the dataset Amazon Prime Pantry 5-core from~\cite{ni2019justifying}. The dataset involves 14,180 users interacted with 4,968 items. A sequence of reviews of items interacted with a specific user is treated as one sample.

We encode movie topics in MovieLens dataset or item reviews in Amazon dataset using the pre-trained encoder of Sentence-BERT~\cite{reimers2019sentence} to serve as the prior to the proposed method.
\vspace{-3mm}
\subsection{Baseline models and ablation study}
We select two models as comparison baselines, Bayesian Personalized Ranking Matrix Factorization (BPRMF)~\citep{bpr} and Self-Attention based Sequential Recommendation (SASRec)~\citep{kang2018self}, which serve as representative recommenders of a Bayesian and a sequential model. Our proposed method, a Bayesian regularizer, is an enhancement to the learning objectives of other recommenders, and not a recommender itself. Therefore, we do not compare to other recommenders.

\textbf{BPRMF} is a specialized algorithm used in recommendation systems, particularly effective for scenarios involving implicit user feedback, like clicks or purchases.  It operates under a Bayesian framework, focusing on providing personalized rankings rather than predicting explicit ratings. The algorithm employs matrix factorization, breaking down a user-item interaction matrix into user and item latent factors, which represent underlying preferences and attributes. 

\textbf{SASRec} leverages self-attention, a mechanism popularized by models like Transformers, to understand and predict user preferences based on their sequence of interactions. This technique recognizes the sequential nature of user behavior, where each action is potentially influenced by previous ones. By applying self-attention, the model can weigh the significance of each interaction in the history of a specific user, allowing it to capture and long-range dependencies within the sequence to achieve personalized recommendation. In addition to using the original SASRec, we tailor it by replacing the item ID encoder with a multilayer perceptron (MLP) to incorporate the item metadata.

We build up the proposed method on two comparison baselines, by adding the graph regularization term in their learning objective. Additionally, we conduct ablation study to evaluate the localization of estimating $\Lambda$ in \eqref{eq:s_emp} as follows. \textbf{BPRMF-ours} and \textbf{SASRec-ours} are based on BPRMF and SASRec, respectively, but with localized estimation of $\Sigma_i$ as presented in \eqref{eq:local}. By contrast, \textbf{BPRMF-ours-global} and \textbf{SASRec-ours-global} are still based on BPRMF and SASRec, but with the global estimator of $\Sigma_i$ (i.e., estimating $\Sigma_i$ using all items in the dataset). Localized estimation of $\Sigma_i$ is expected to have more fine-grained variance estimation and less computational burden, and therefore produces enhanced performance.
\vspace{-3mm}
\subsection{Evaluation metrics}
\textbf{Normalized Discounted Cumulative Gain (NDCG)} compares rankings to an ideal order where all relevant items are at the top of the list. NDCG at $N$ is determined by dividing the Discounted Cumulative Gain (DCG) by the ideal DCG representing a perfect ranking. DCG measures the total item relevance in a list with a discount that helps address the diminishing value of items further down the list. Higher NDCG indicates better recommendation performance.

\textbf{Hit Ratio (HR)} is the fraction of users for which the correct answer is included in the recommendation list. Higher HR indicates better recommendation performance.
\vspace{-3mm}
\subsection{Insights from experimental results}

We evaluate the proposed approach against comparison baselines on two real-world datasets according to: (1) the overall performance on general recommendation task (Table~\ref{tab:eval}); (2) the performance on cold-start item recommendation (Table~\ref{tab:eval_cs}); (3) the effect of the localized and the global variance estimator on cold-start item recommendation (Table~\ref{tab:eval_cs}) and (4) the effect of the hyperparameter that penalizes the graph regularization (Figure~\ref{fig:ndcg} and Figure~\ref{fig:hr}). From the experimental results, we can conclude that the proposed approach demonstrates superior performance. Additionally, we have the following observations.
\begin{table*}[hbt!]
 \caption{Overall performance of proposed method by evaluating on users in testing set ($\rho=1$). Compare SASRec-ours with SASRec, BPRMF-ours with BPRMF in two datasets at all three metrics.}
\vspace{-3mm}
\centering
\begin{adjustbox}{width=\textwidth,center}
\begin{tabular}{c c cccccc c cccccc} 
\hline
\multirow{2}{*}{Methods} && \multicolumn{6}{c}{MovieLens 25M} && \multicolumn{6}{c}{Amazon} \\\cline{3-8} \cline{10-15}
&& NDCG@10 & NDCG@20 & NDCG@40 & HR@10 & HR@20 & HR@40 && NDCG@10 & NDCG@20 & NDCG@40 & HR@10 & HR@20 & HR@40 \\ \hline
BPRMF &&0.0100 &0.0155 &0.0345 &0.0206 &0.0410 &0.1046 &&0.2515 &0.2871 &0.3015 &0.8905 &0.9898 &0.9996 \\
BPRMF-ours &&0.0201 &0.0312 &0.0743 &0.0287 &0.0565 &0.1324 &&0.3855 &0.3932 &0.3981 &0.9474 &0.9903 & 0.9996 \\
\hline
SASRec &&0.0199 &0.0361 &0.1011 &0.0434 &0.1099 &0.4502 && 0.3734 &0.3839 &0.3848 &0.9594 &0.9954 &0.9996 \\
SASRec-ours &&\textbf{0.0309} &\textbf{0.0479} &\textbf{0.1198} &\textbf{0.0467} &\textbf{0.1167} &\textbf{0.4683} && \textbf{0.4311} &\textbf{0.4360} &\textbf{0.4363} &\textbf{0.9802} &\textbf{0.9986} &\textbf{1.0000} \\

\hline
\end{tabular}
\end{adjustbox}
\label{tab:eval}
\end{table*}
\begin{table*}[hbt!]
 \caption{Overall performance on cold-start items of proposed method by evaluating on users that interact with at least one cold-start item. Compare SASRec-BayesRec with SASRec, LightGCN-BayesRec with LightGCN in two datasets at all three metrics.}
\vspace{-3mm}
\centering
\begin{adjustbox}{width=\textwidth,center}
\begin{tabular}{c c cccccc c cccccc} 
\hline
\multirow{2}{*}{Methods} && \multicolumn{6}{c}{MovieLens 25M} && \multicolumn{6}{c}{Amazon} \\\cline{3-8} \cline{10-15}
&& NDCG@10 & NDCG@20 & NDCG@40 & HR@10 & HR@20 & HR@40 && NDCG@10 & NDCG@20 & NDCG@40 & HR@10 & HR@20 & HR@40 \\ \hline
BPRMF &&0.0124 &0.0220 &0.0585 &0.0345 &0.0813 &0.1824 && 0.1564&0.2455 &0.2612 &0.5552 &0.8565 &0.9605 \\
BPRMF-ours-global && 0.0145 &0.0236 &0.0664 &0.0412 &0.0909 &0.2034 &&0.2012 &0.3105 &0.3263 &0.6718 &0.8768 &0.9889 \\
BPRMF-ours &&0.0186 &0.0285 &0.0998 &0.0640 &0.1515 &0.2755 && 0.3750&0.3842 &0.4010 &0.8650 &0.9645 &0.9964 \\
\hline
SASRec &&0.0438 &0.0668 &0.1129 &0.0875 &0.1798 &0.4060&& 0.2447 &0.2975 &0.3170 &0.6687 &0.8755 &0.9695 \\
SASRec-ours-global &&0.0580 &0.0831 &0.1298 &0.0918 &0.1629 &0.4030&&0.4100 &0.4165 &0.4303 &0.9061 &0.9655 &0.9915 \\
SASRec-ours &&\textbf{0.0723} &\textbf{0.0898} &\textbf{0.1341} &\textbf{0.0979} &\textbf{0.1887} &\textbf{0.4072} && \textbf{0.4791} &\textbf{0.4963} &\textbf{0.5007} &\textbf{0.9090} &\textbf{0.9757} &\textbf{0.9964} \\
\hline
\end{tabular}
\end{adjustbox}
\label{tab:eval_cs}
\vspace{-5mm}
\end{table*}

% \todob{Subsubsections are an overkill for an $8$-page paper.}

% \subsubsection{The proposed method achieves enhanced overall performance on recommendation}

% \todob{Break the long paragraph below into shorter paragraphs along the points that you want to make:

% * SASRec beats BPRMF. This should be the last point.

% * We improve SASRec.

% * We improve BPRMF.}

\textbf{The proposed method achieves enhanced overall performance on recommendation}. Although the proposed method is designed to recommend cold-start items, we still evaluate its performance on general recommendation (i.e., recommendation on a mixture of cold-start and general items). As illustrated in Table.~\ref{tab:eval}, we find the sequential model (i.e., SASRec) outperforms the Bayesian model with matrix factorization (i.e., BPRMF). On MovieLens dataset, SASRec outperforms BPRMF by 161.83$\%$ regarding NDCG on average, and by 263.12$\%$ regarding HR on average. On Amazon dataset, SASRec outperforms BPRMF by 55.15$\%$ regarding NDCG on average, and by 2.59$\%$ regarding HR on average. Besides, we note that the performance of the proposed methodologies, BPRMF-ours and SASRec-ours, exhibits a marked enhancement in comparison to the original versions of these methods. For instance, the NDCG and HR of BPRMF-ours are 109.33$\%$ and 30.93$\%$ larger than BPRMF, respectively, in MovieLens dataset. In Amazon dataset, NDCG and HR of BPRMF-ours are 40.08$\%$ and 1.99$\%$ larger that BPRMF, respectively. We find similar pattern on SASRec, where NDCG and HR of SASRec-ours are 26.42$\%$ and 4.67$\%$ larger than SASRec in MovieLens dataset, respectively. In Amazon dataset, NDCG and HR of SASRec-ours are 14.12$\%$ and 0.83$\%$ larger than SASRec, respectively. The superior performance of the proposed method potentially results from the cold-start items contained in the dataset. The proposed method is able to well learn the behavior of those cold-start items, which cannot be captured by conventional recommenders.

% \subsubsection{The proposed method achieves superior performance on cold-start item recommendation}
\textbf{The proposed method achieves superior performance on cold-start item recommendation}. To evaluate the performance of the proposed method on recommending cold-start items, we extract users that interacted with at least one cold-start items in both MovieLens and Amazon datasets. Then we apply the proposed approach and comparison baselines to those subsets. As shown in Table~\ref{tab:eval_cs}, the proposed approach is consistently superior. For instance, in MovieLens dataset, NDCG and HR of BPRMF-ours are larger than BPRMF by 58.13$\%$ and 64.65$\%$ on average, respectively. In Amazon dataset, NDCG and HR of BPRMF-ours are larger than BPRMF by 74.97$\%$ and 19.13$\%$ on average, respectively. Similar pattern is observed on the sequential recommender. Specifically, SASRec-ours is superior than SASRec by 32.53$\%$ and 71.80$\%$ on average regarding NDCG in the MovieLens and the Amazon dataset, respectively. In terms of HR, SASRec-ours outperforms SASRec by 3.04$\%$ and 14.62$\%$ on average, respectively. The superior performance of the proposed method by leveraging graph regularization to incorporate the prior knowledge based on item metadata potentially attributes to well-captured item similarity by the kernel (i.e., \eqref{eq:s_emp}). The kernel enhances the item embeddings learned by the recommender that recommends cold-start items in the similar manner as recommending similar items.

\textbf{Localized covariance estimator has improved performance against the global estimation}. In addition to evaluating the performance on recommendation, we also assess the effect of covariance estimation in \eqref{eq:local}. The intuition of estimating localized covariance using surrounding $K$ nearest items is to save the complexity of the estimation process from $O(N)$ to $O(K)$. Also, the localized estimation potentially achieves fine-grained estimation of the covariance, but at the risk of poor generalization to other items during the inference phase. We compare the proposed approach using localized estimation with $K=\sqrt{N}$ (i.e., BPRMF-ours and SASRec-ours) against those using glocal estimation of the covariances using all items in the dataset (i.e., BPRMF-ours-global and SASRec-ours-global). As shown in Table~\ref{tab:eval_cs}, BPRMF with the locally estimated covariance achieves 40.57$\%$ and 38.45$\%$ higher NDCG on average than the one with globally estimated covariance, in MovieLens and Amazon datasets, respectively. Additionally, in terms of HR, the localized estimator outperforms the global estimator by 5.49$\%$ and 0.63$\%$ on average in MovieLens and Amazon, respectively. This indicates that fine-grained estimation on covariance plays a significant role in the accuracy of recommending cold-start items. Estimating the covariance using all items might not provide informative insights in learning representation of similarly items.

% \subsubsection{Strength of penalty on graph regularization is crucial}
\textbf{Strength of penalty on graph regularization is crucial}. Last but not the least, we evaluate the impact of hyperparameter (i.e., $\rho$ in \eqref{eq:obj_reg_local}) in the recommendation performance. The hyparameter intuitively measures the strength of the graph regularization. When it is too small, then the similarity captured by the prior may not help learning item representation. When the hyperparameter is too large, then the graph regularization forces similar items to have the same representation, leading to oversmoothing issues. Figure~\ref{fig:ndcg} and Figure~\ref{fig:hr} measure the relative NDCG and HR compared to the baseline (e.g., $\rho=0$), respectively. As shown in Figure~\ref{fig:ndcg}, when $\rho$ increases, the NDCG will first increases and then drops when evaluating on MovieLens dataset, for both general and cold-start item recommendation. In Amazon dataset, the NDCG will increase when $\rho$ increases. This might be due to more diverse items contained in Amazon dataset, causing the item representation learning assisted by similar items much challenging. In terms of HR (Figure~\ref{fig:hr}), in MovieLens dataset, we observe similar pattern as the NDCG of recommending general items will increase and then drop when $\rho$ increases. The NDCG will drop in general when $\rho$ increases when recommending cold-start items, while it is still higher than the baseline (i.e., $\rho=0$). In Amazon dataset, when $\rho$ increases, HR will increase when recommending cold-start items and stay slightly higher than the baseline (i.e., $\rho=0$) when recommending general items.

\vspace{-3mm}
\section{Conclusion}
We propose a novel approach for cold-start item recommendation that leverages LMs to inject prior knowledge of items and integrates a Bayesian regularizer during the training process of the RecSys. Our experimental results demonstrate enhanced performance of recommending cold-start items of the proposed approach compared to baselines. Particularly, our method can be adapted to the learning objective of any sequential or collaborative filtering-based recommenders, as long as item metadata is available.

We develop our method from publicly available MovieLens 25M\footnote{https://grouplens.org/datasets/movielens/25m/} and Amazon Prime Pantry 5-core~\cite{ni2019justifying} datasets. It is important to note that, like other recommenders, our implementation will likely reflect the socioeconomic and entity biases inherent in datasets that we use. Additionally, although our method is designed for cold-start item recommendation, we are not able to control the item that user would recommend, which may contain improper contents.

\newpage

% References
\bibliography{References}

\onecolumn
\appendix
\section{Reproducibility}
Datasets and code to reproduce results of the paper is in \url{https://github.com/as68578688/LLM4RecSysColdStart}. Code is proprietary and will be released soon upon approval.

\end{document}